# All-optical measurement-device-free feedforward enabling ultra-fast quantum information processing

TAICHI YAMASHIMA,[1,2] TAKAHIRO KASHIWAZAKI,[2] TAKUMI SUZUKI,[1] RAJVEER NEHRA,[1,3,4,5] TOMOHIRO NAKAMURA,[1] ASUKA INOUE,[2] TAKESHI UMEKI,[2] KAN TAKASE,[1,6] WARIT ASAVANANT,[1,6] MAMORU ENDO[1,6] AND AKIRA FURUSAWA[1,6,*]

[1]Department of Applied Physics, School of Engineering, The University of Tokyo, 7-3-1 Hongo, Bunkyo-ku, Tokyo 113-8656, Japan
[2]NTT Device Technology Labs, NTT Corporation, 3-1 Morinosato Wakamiya, Atsugi, Kanagawa 243-0198, Japan
[3]Department of Electrical and Computer Engineering, University of Massachusetts Amherst, Amherst, Massachusetts 01003, USA
[4]Department of Physics, University of Massachusetts Amherst, Amherst, Massachusetts 01003, USA
[5]College of Information and Computer Science, University of Massachusetts Amherst, Amherst, Massachusetts 01003, USA
[6]Optical Quantum Computing Research Team, RIKEN Center for Quantum Computing, 2-1 Hirosawa, Wako, Saitama 351-0198, Japan
*akiraf@ap.t.u-tokyo.ac.jp

**Abstract:** Utilizing feedforward to perform adaptive quantum operations on one entangled state according to the measurement result of the other state enables measurement-based quantum information processing (QIP). Until now, the bandwidth of feedforward in optical QIP has been limited to around 100 MHz by measurement with electronics. A potential alternative is the utilization of an optical parametric amplifier (OPA). This optical device eliminates the need for electronic measuring devices and enables all-optical broadband feedforward. In this paper, we demonstrate a variable squeezing gate with an operation bandwidth of 1.3 THz by all-optical measurement-device-free feedforward. We utilize a periodically poled lithium niobate waveguide as a broadband OPA and perform continuous phase locking in our optical system. Experimental results demonstrate that our all-optical QIP operates at a THz clock frequency, representing a major step toward the realization of an ultra-fast quantum computer.

## 1. Introduction

Quantum information processing (QIP) is expected to solve certain computational problems much faster than conventional processing technique. Research on the development of a quantum computer is underway in various physical systems, such as superconducting circuits [1], trapped-ion systems [2], and so on. Among these, optical circuits are expected to enable the construction of a scalable quantum computer following measurement-based quantum computing (MBQC) model with time-domain multiplexing technique [3–5]. The operation of the optical quantum computer can also be ultra-fast thanks to a high carrier frequency of hundreds of THz and a wide range of sidebands [6]. Moreover, utilizing continuous variables of light is considered promising because MBQC or QIP becomes deterministic [7]. This approach has been intensively studied following the achievement of deterministic quantum teleportation [7]. Recently, a large-scale entangled state required for MBQC has been generated in the time domain [8–10] , and squeezed vacuum, which is the resource of entanglement, has been generated over the bandwidth of 6 THz with a high squeezing level [11]. These results suggest that we are now able to prepare a broadband entangled state within ultra-short wave packets in the time domain for scalable, ultra-fast and deterministic QIP.

According to the MBQC model, we can perform ultra-fast QIP by a series of measurements and feedforward on the broadband entangled state prepared in ultra-short wave packets. Even though each measurement on one entangled state is non-unitary, the whole process of the QIP becomes unitary by feedforward, which is performing adaptive quantum operations on the other entangled state according to the measurement result. In addition to this fact that the feedforward enables MBQC, feedforward according to detected errors also enables quantum error correction by utilizing quantum entanglement [12]. Thus, performing measurements and feedforward is a crucial technique and must also be broadband for ultra-fast QIP.

However, despite the broadband nature of light, the bandwidth of quantum operation using continuous variables of light is still limited. This is due to the conventional "opto-electro-optical" feedforward where the intermediate electrical signal intrinsically limits the operation bandwidth of the whole process. In the conventional method, one entangled state is measured by a homodyne detector with photodiodes and converted into a classical electrical signal, the signal is sent to an electro-optical device to generate a coherent state in the sideband, and the state is then used for adaptive displacement operation on the other of the entangled state. Since current electronic circuits for measurements can only handle signals up to around 100 GHz, the shape of wave packets must be rounded accordingly, which leads to slower operation than would have been possible in the optical region. In addition to this principal limitation, the use of photodiodes with high quantum efficiency and thick absorbing layers for measurements further limits the operation bandwidth of quantum operation to around 100 MHz at best [13–15].

In this work, we demonstrate an all-optical measurement-device-free feedforward that takes full advantage of the broadband nature of optics and report the results of experiments showing that an all-optical squeezing gate operates with an operation bandwidth of 1.3 THz. In the all-optical method [16] we use, we replace the homodyne detector with an optical parametric amplifier (OPA). The OPA amplifies one of the quadrature variables, which are continuous variables of light, while de-amplifying the other. When its gain is large enough, the output of the OPA can be regarded as carrying classical information of only one quadrature [16]. This is the same as what a homodyne detector does, except that the information is an optical signal. Thus, the feedforward process is free from measurement devices and electronics. Up to now, while all-optical quantum operations using Rb gas as an OPA have been reported [17,18], their operation bandwidth is around 10 MHz, which is narrower than that of conventional feedforward. In our experiment, we utilize a periodically poled lithium niobate (PPLN) waveguide[11,19–21] as an OPA. We do not use the OPA only for broadband detection (as in [22–24]) but also for all-optical feedforward to achieve QIP. The operation bandwidth of an all-optical QIP is greatly extended thanks to the broad bandwidth of the PPLN waveguide, as we achieved the operation bandwidth of 1.3 THz. In addition, phase locking in our optical system is performed continuously. This implies that our QIP operates at a THz clock frequency, which is improved by around four orders of magnitude. This is a major step towards the realization of an ultra-fast quantum computer.

(a)

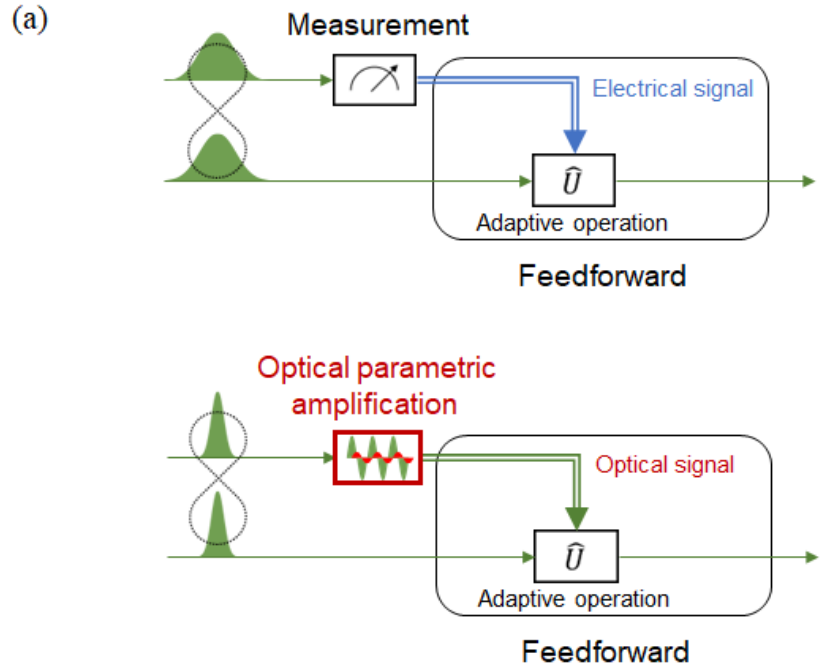

(b)

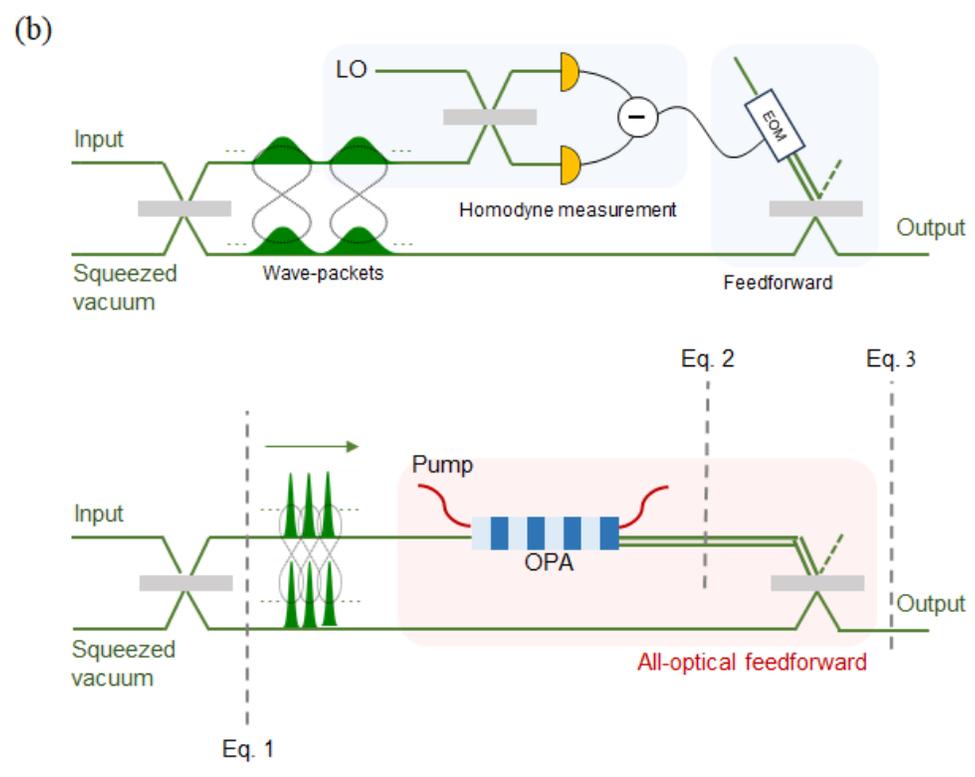

**Fig. 1.** Conceptual diagrams of quantum information processing (QIP), where one entangled state is measured and the result is used for adaptive operation on the other. (a) Conventional "opto-electro-optical" feedforward, where the intermediate electrical signal intrinsically limits the operation bandwidth of the whole process. (b) All-optical feedforward, where an OPA replaces a detector and removes the bandwidth limitation by electronics. (c) Universal squeezing gate with the conventional feedforward. The use of photodiodes with high quantum efficiency and thick absorbing layers for measurements strictly limits the operation bandwidth. As a result, wave packets are inevitably defined longer in the time domain. (d) Universal squeezing gate with an all-optical feedforward. As this method is free from photoelectric conversion and electronics, the operation bandwidth can extend up to the THz range, which is the inherent nature of optics. Therefore, ultra-short wave packets can be defined in the time domain.

## 2. Method

In the ideal scenario for CV-QIP, the EPR state has infinite noise in both conjugate continuous variables, quadrature $\hat{x}$ and $\hat{p}$. In quantum teleportation, there is no destruction of quantum states in the Bell measurement because their information is obscured by the infinite noise. The randomness caused by this noise is then counteracted by feedforward. This is how the whole process becomes a unitary and deterministic operation.

In our experiment, we built the setup shown in Fig.1(d), which constitutes a universal squeezing gate[25,26], a type of QIP. In this gate operation, only one of the quadratures is masked by the noise from the squeezed vacuum, which is the resource of the EPR state. Feedforward is also applied only to this quadrature. In this paper, we demonstrate all-optical feedforward and apply the squeezing gate operation to the input vacuum, confirming that the noise from the squeezed vacuum in the quadrature $\hat{p}$ is canceled out while the quadrature $\hat{x}$ remains as it was before feedforward.

Figure 1(c) shows a universal squeezing gate with the conventional feedforward. The electronics limit the operation bandwidth no matter how broadly entangled states can be prepared. In the time domain, wave packets are inevitably defined longer. Figure 1(d) shows a universal squeezing gate with an all-optical feedforward. The broadband OPA replaces the homodyne detector and frees us from electronics, resulting in the operation bandwidth up to the THz region. Ultra-short wave packets can be defined in the time domain.

We now describe the theory of the universal squeezing gate with all-optical feedforward. If we write the annihilation operators of the input and ancillary state as $\hat{a}_{\text{in}}$ and $\hat{a}_{\text{anc}}$, then the output of the beam splitter with transmission T can be written as

$$\hat{a}'_{\text{in}} = \sqrt{1-T}\hat{a}_{\text{in}} - \sqrt{T}\hat{a}_{\text{anc}},$$
$$\hat{a}'_{\text{anc}} = \sqrt{T}\hat{a}_{\text{in}} + \sqrt{1-T}\hat{a}_{\text{anc}}, \quad (1)$$

where the annihilation operator $\hat{a}$ is represented by quadrature operator $\hat{x}$ and $\hat{p}$ as $\hat{a} = (\hat{x} + i\hat{p})/\sqrt{2}$ assuming $\hbar = 1$. Here, the ancillary state is a squeezed vacuum with the squeezing parameter $r$, that is, $\hat{a}_{\text{anc}} = (e^{-r}\hat{x}_{\text{vac}} + ie^{r}\hat{p}_{\text{vac}})/\sqrt{2}$, where the subscript "vac" refers to a vacuum. One of the outputs then passes through an OPA with parametric gain $G$ and its output is represented as

$$\hat{a}'_{\text{in}} \rightarrow \hat{a}'_{\text{in},G} = \frac{1}{\sqrt{2G}} \hat{x}'_{\text{in}} + i\sqrt{\frac{G}{2}} \hat{p}'_{\text{in}}. \tag{2}$$

When the parametric gain $G$ is large, this output can be regarded as carrying only classical information about quadrature $\hat{p}$. The feedforward is completed by applying the displacement only to quadrature $\hat{p}$, that is $\hat{x}'_{\text{anc}} \rightarrow \hat{x}''_{\text{anc}} = \hat{x}'_{\text{anc}}$ and $\hat{p}''_{\text{anc}} = \hat{p}'_{\text{anc}} + g\hat{p}'_{\text{in},G}$ where $g$ is an experimentally tunable factor. By choosing $g = \sqrt{(1-T)/GT}$, we can cancel out the noise from the squeezed vacuum and acquire the output of the universal squeezing gate as follows:

$$\hat{x}''_{\text{anc}} = \sqrt{T}\hat{x}_{\text{in}} + \sqrt{1-T}e^{-r}\hat{x}_{\text{vac}}$$
$$\hat{p}''_{\text{anc}} = \frac{1}{\sqrt{T}}\hat{p}_{\text{in}} \tag{3}$$

This feedforward can be achieved by interfering the beam output from the OPA, which has only the information of the quadrature $\hat{p}$, with the target beam using a single beam splitter. The gain $g$ is tuned by adjusting the gain of the OPA or attenuating the power of its output. In our experiment, we use the latter method. This operation serves as an ideal squeezing gate in the limit of infinite squeezing, $r \rightarrow \infty$, while the output has an additional term due to the finite squeezing level. In addition, this output state is degraded by optical losses in real experiments.

Unlike the conventional method, this all-optical feedforward is free from any measurement device or non-unitarity of measurements because the OPA intrinsically works as a unitary and reversible operation. There is irreversibility at the variable attenuator or the beam splitter for displacement because light is partially rejected [16]. By performing feedforward, the whole process in principle becomes a unitary operation.

To measure the squeezing level of this output state and ensure that the feedforward is successful over the THz bandwidth, we use an intensity measurement with an optical spectrum analyzer rather than an amplitude measurement with a homodyne detector. This is because amplitude measurement using electronics is so slow that it cannot measure the squeezing level over the THz bandwidth. To measure the variance of the quadrature from its intensity, we use a vacuum as the input state, which has zero average amplitude.

### 3. Measurement Setup

Figure 2 shows the setup for this experiment. First, we explain the optical setup. A continuous wave laser at the wavelength of 1545 nm and 773 nm is output from the fiber-laser. The carrier light (1545 nm) is divided into two beams utilized as probe beams for phase locking. One is phase-modulated at 2 MHz by an electro-optic modulator (EOM1) (MPX-LN-0.1-00-P-P-FA-FA, iXblue Photonics) and injected through the input path of the squeezing gate. The other is frequency-shifted by +1.06 MHz by a pair of acoustic-optic modulators (AOMs) (SGTF-40-1550-1P, Chongqing Smart Science &Technology Development) and injected from the ancillary path. Another EOM (EOM2) is inserted in its optical path as an actuator for phase locking. The pump light (773 nm) is divided into three beams that are injected into three OPA modules. A fiber stretcher is inserted in the path of the second pump beam as an actuator for phase locking. We explain the setup of the squeezing gate below. The input state of the squeezing gate is a vacuum as explained earlier. The squeezed vacuum, i.e., the ancillary state, is generated in the first OPA module (OPA1). OPA1 is fabricated with a dicing method[19,20] to generate as highly squeezed vacuum as possible. These states pass through a variable beam splitter that consists of two polarizing beam splitters and a half-wave plate. We utilize a variable beam splitter to ensure that we can perform all-optical feedforward for various splitting ratios. One of the outputs is injected into the second OPA module (OPA2), where its quadrature $\hat{p}$ is amplified and $\hat{x}$ is de-amplified. OPA2 is fabricated with a dry-etching method [19,21,27] to earn as much gain as possible. An additional fiber (PMDCFA2, Thorlabs) is connected to compensate for the dispersion induced by the output fiber of OPA2. The gain of OPA2 is fixed, and its output power is appropriately attenuated to exactly cancel out the entangled noise in quadrature $\hat{p}$. This attenuation is allowed because the output of the OPA2 carries only classical information. The other output of the variable beam splitter passes through a variable delay line so that both outputs propagate the same distance to the 99:1 beam splitter for displacement. To cancel out the noise from the squeezed vacuum simultaneously over the THz bandwidth, it is necessary to adjust the optical path length by an order of $(3 \cdot 10^8/10^{12})/360 \sim 1~\mu$m. Here, the phase is targeted to match with an accuracy of $1°$. The optical path is precisely adjusted by a variable delay line (VDL) made with a manual translation stage with two mirrors on it. Since the intensity of the output is too weak as it is, we use a third OPA module (OPA3) to amplify its quadrature and then measure its intensity, following the method in [22,23]. The intensity of light output from OPA3 is measured by an optical spectrum analyzer (AD6370D, YOKOGAWA). The gain of OPA3 is precisely controlled by setting the feedback loop with reference to the pump power output from OPA3 so that the values of the squeezing and anti-squeezing levels to be measured are reliable.

Next, we explain the electrical setup for phase locking. In this experiment, there are five phase locking points: interference between the pump light and the carrier light at the three OPA modules, interference between the two probe beams, and interference at the 99:1 beam splitter. The most difficult part is the interference at the 99:1 beam splitter, as the interference signal is smaller than the non-interference signals due to the large bias in the splitting ratio. We perform this phase locking by using two types of probe beams, one phase-modulated by EOM and the other frequency-shifted by AOM. The 1% optical power is detected at each phase locking point, and each signal is multiplied by a mixer with an electrical sine wave from each local oscillator. Phase locking is then achieved by sending those signals to the actuators via servo amplifiers. The actuators used are EOMs, a fiber stretcher (FST), and piezoelectric transducers (PZTs) attached to mirrors. While the sample-and-hold method [14,15] makes phase locking relatively easy, the clock frequency of the QIP would drop even with a high operation bandwidth because we cannot perform the quantum operation while sampling. In this experiment, we do not use the sample-and-hold method, and the phase locking is carried out continuously in time. Thus, the broad operation bandwidth directly leads to a high clock frequency.

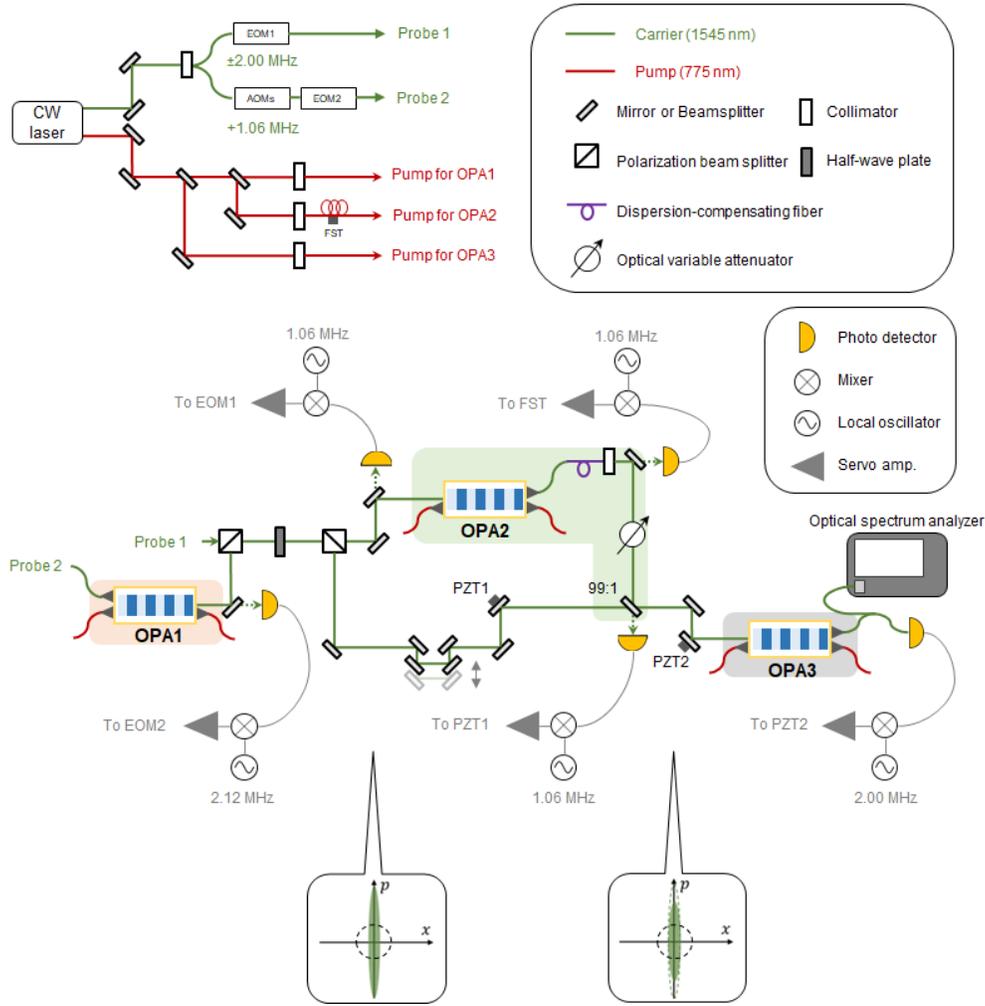

**Fig. 2.** Experimental setup. Green(red) lines represent carrier light (pump light). Straight (curved) lines represent light traveling in free space (in optical fibers). We use three OPA modules for different purposes: generation of squeezed vacuum (highlighted in red), all-optical feedforward (green) and intensity measurement of the output (gray). The two figures in balloons represent the variance of quadrature before and after feedforward.

In this experimental system, OPA1 module has an optical loss of approximately 4%, which is the effective loss in the PPLN waveguide fabricated with a dicing method [19]. As for the output, the extra coupling loss is avoided by removing the output fiber of the module and allowing the output beam to propagate in free space. The variable beam splitter has a loss of approximately 7%. For the upper arm of the output of the variable beam splitter, there is 1% loss by tapping for phase locking, 11% coupling loss at OPA2, and 5% effective propagation loss in OPA2. The effective propagation loss is calculated from the transmittance of the PPLN waveguide and its gain of 28.4 dB (See the Appendix). The input fiber of OPA2 is removed.

The length of the output fiber is approximately 1 m, which corresponds to 1.5 m in free space. For the lower arm, there is a total 8% scattering loss at the mirrors for making the optical path longer and at lenses for aligning the spatial mode of the OPA2 output with that of the collimator. At the displacement, a 99:1 beam splitter is used to approximately perform the feedforward described in the Method section. At the intensity measurement, there is 15% coupling loss at OPA3 and 7% effective propagation loss in OPA3 whose gain is 20.7 dB. The input fiber of OPA3 is also removed.
.

**Table 1.** Purpose and performance of three OPAs. Total loss consists of coupling loss and propagation loss. The loss in OPA1 does not include coupling loss because its input is vacuum.

|      | Purpose                       | Total loss           | Gain    |
|------|-------------------------------|----------------------|---------|
| OPA1 | Generation of squeezed vacuum | 4% (no coupling loss) |         |
| OPA2 | All-optical feedforward       | 15 %                 | 28.4 dB |
| OPA3 | Intensity measurement         | 21 %                 | 20.7 dB |

## 4. Results

The squeezing level of the ancilla generated by OPA1 is measured as $3.6\pm0.2$ dB and the anti-squeezing level as $9.3\pm0.2$ dB in this measurement setup. The loss calculated from these squeezing and anti-squeezing values is 39%, which is consistent with the actual loss $(1-0.96\cdot0.93\cdot0.92\cdot0.99\cdot0.79) \sim 36\%$. The gain of OPA2 is 28.4 dB, which indicates that its output is carrying classical information. By tuning the gain of feedforward and the delay line, we first confirm the cancellation level [14,15] of more than 30 dB by homodyne measurement in the sideband up to 40 GHz. We insert another EOM (MPZ-LN-40-00-P-P-FA-FA, iXblue Photonics) in the path of Probe2 only when this cancellation measurement is conducted. Finally, fine adjustments are made while monitoring the display on the optical spectrum analyzer. The gain of OPA3 for intensity measurement is 20.7 dB, which is sufficient to measure the squeezing level in this case [28]. Also, the shotnoise fluctuation over five minutes is suppressed to within just $\pm0.2$ dB thanks to the feedback control.

Figure 3(a) shows the results by the optical spectrum analyzer for various values of $T$. We can see here that the feedforward process cancels out the noise in the quadrature $\hat{p}$ in the sideband over the 1-THz bandwidth. At the same time, there is almost no difference in the quadrature $\hat{x}$ between before and after feedforward. The values for shaded areas (0 to $\pm0.1$ THz) are not used in the analysis because the power of carrier light used for phase locking is dominant in this bandwidth due to the finite resolution bandwidth of OSA (0.02 THz). In future experiments, this carrier frequency power can be eliminated by chopping. Figure 3(b) shows the optical power normalized by shot noise level in the single sideband from 0.01 to 5 THz. On the basis of this result, we use data up to $\pm1.3$ THz where the squeezing and anti-squeezing level seem flat (highlighted in light red). Figure 3(c) shows the variance of quadrature before and after feedforward. The variance is calculated from the average of optical power over both sidebands from $-1.3$ to $-0.1$ THz and from 0.1 to 1.3 THz. To confirm whether this operation is successful, we calculate the product of the squeezing and anti-squeezing level of the quadrature. These values should approach 1 after the feedforward cancels out the excess noise in the quadrature $\hat{p}$. The results show that the values of the product decreased from 3.7 to 1.6 for $T = 0.62$, from 3.5 to 1.4 for $T = 0.50$, from 3.0 to 1.4 for $T = 0.40$, and from 2.7 to 1.2 for $T = 0.30$.

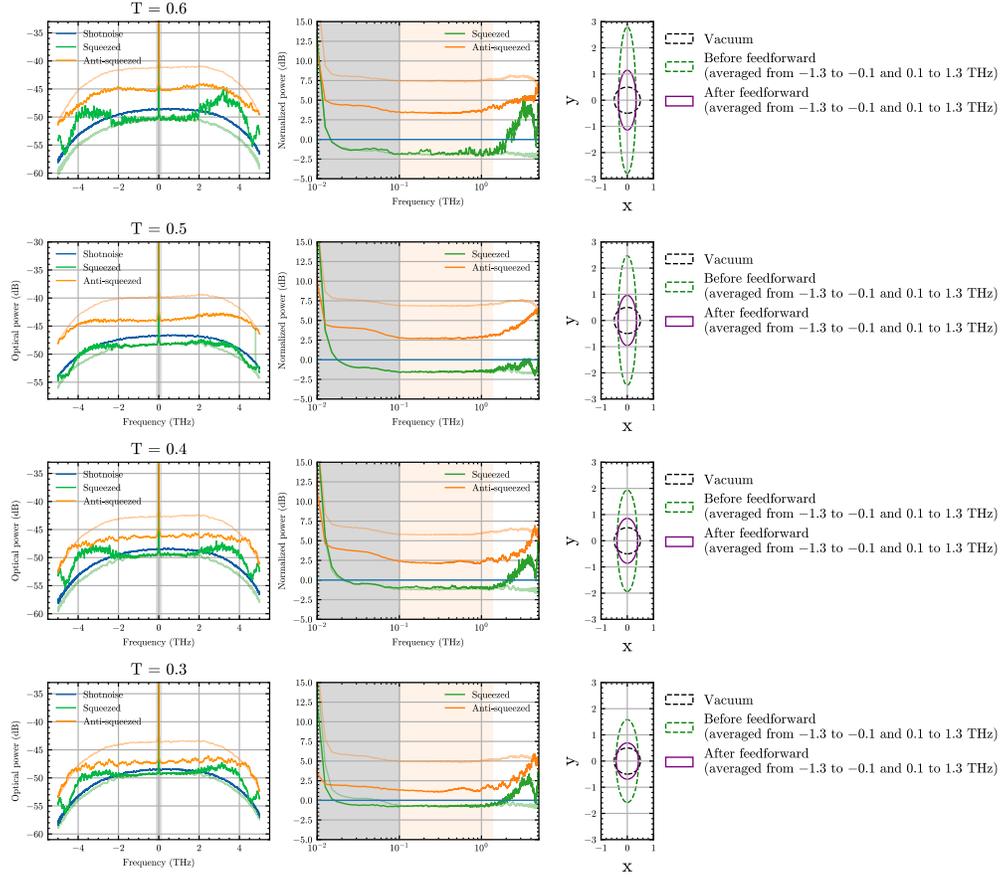

**Fig. 3.** Results of the experiment. (a) Intensity measurement by optical spectrum analyzer (OSA). The blue line represents shot noise level. The orange (green) lines represent anti-squeezing (squeezing) level. Lighter lines show each level before feedforward. The values in sideband from 0 to $\pm 0.1$ THz are not considered in the data analysis because the power of carrier light used for phase locking is dominant in this bandwidth due to the finite resolution width of OSA. This bandwidth is shaded in the graphs. (b) Optical power in the single sideband from 0.01 to 5 THz. The power is normalized by shot noise and the horizontal axis is logarithmic scale. The bandwidths used in the analysis are highlighted in light red. (c) Variance of quadrature calculated by averaging the optical power from $-1.3$ to 1.3 THz (purple line) except the bandwidth from $-0.1$ to 0.1 THz. The black (green) line represents the variance of vacuum (output without feedforward). Each ellipse is plotted with the reference to the values of anti-squeezing and squeezing level.

Figure 4 shows the results with various values of transmittance $T$. The error bar is set to $\pm 0.2$ dB due to fluctuation of the shotnoise level. The light-colored curves are theoretical curves. The theoretical values are calculated as below by modifying Eq.3 considering experimental optical losses in coupling and propagation at each OPA.

$$S_{\text{out}}^{+} = (1 - l_3)\left(\frac{1}{T} + \frac{1-T}{T}\frac{l_2}{1-l_2}\right) + l_3$$

$$S_{\text{out}}^{-} = (1 - l_3)\{T + (1-T)S_{\text{anc}}^{-}\} + l_3$$

(4)

$S_{\text{anc,out}}^{\pm}$ denote the squeezing (−) and anti-squeezing (+) level of ancilla and the output of squeezing gate. $l_2$ and $l_3$ are the total optical loss in OPA2 and OPA3 respectively. The losses by 99:1 beam splitters used for phase locking or displacement are ignored above for simplicity. The measured values closely follow the theoretical curves.

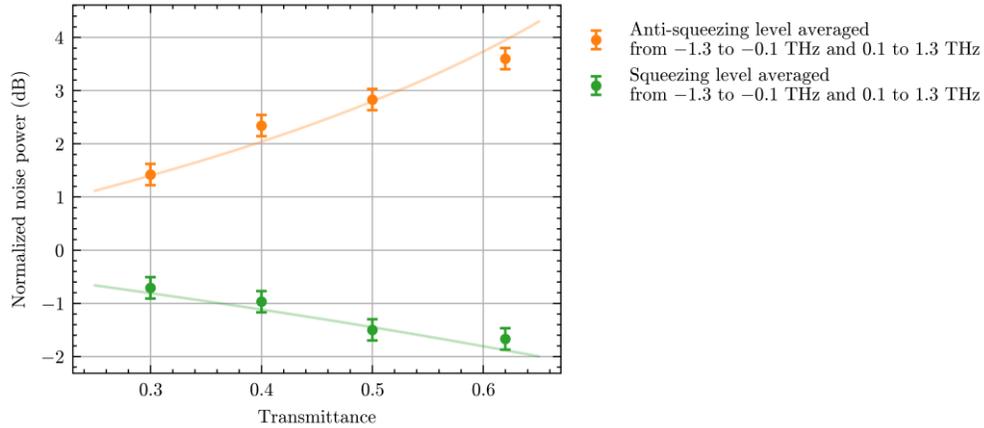

**Fig. 4.** Squeezing and anti-squeezing level at each transmittance $T$. The curves represent theoretical values calculated from Eq.4 considering optical losses in the experiment. The variance is calculated from the average of optical power over both sidebands from −1.3 THz to −0.1 THz and from 0.1 THz to 1.3 THz.

## 5. Discussion

In this work, we have presented our implementation of all-optical measurement-device-free feedforward and demonstrated a variable squeezing gate. The results in Fig.3 confirm that the entangled noise in the quadrature $\hat{p}$ is canceled out while the quadrature $\hat{x}$ remains as it was before feedforward over the 1.3-THz bandwidth. We can also confirm that the product of the squeezing and anti-squeezing level approaches unity by feedforward. In addition, the results in Fig.4 confirm that our squeezing gate works almost exactly as Eq.4 suggests with various values of $T$. From these findings, we conclude that we have achieved an all-optical variable squeezing gate with an operation bandwidth of 1.3 THz. This is the first implementation of QIP with THz operation bandwidth.

In our experiments, the operation bandwidth was limited by the dispersion in an output fiber connected with the OPA2 module, and we relaxed this limitation up to around 1 THz by using a dispersion-compensating fiber. Therefore, above around 2 THz, the interference phase is not aligned and thus feedforward is not performed correctly. For broader operation bandwidth, the use of a waveshaper could be considered in future experiments. Since the OPA2 emission has only classical information, the optical loss due to such devices is not a problem. In principle, the bandwidth of the feedforward can be broadened to that of the OPA1 and OPA2, whose operation bandwidths are almost equivalent. Some ripples are observed in the anti-squeezing in Fig.3(a), and we confirmed that this ripple increases as the power of the probe beams used for phase locking increases. We then minimized the power to keep this effect as small as possible. One possibility is that the probe beam for phase locking is causing a cascade of difference frequency generation or sum frequency generation at OPA2 and OPA3.

To improve the accuracy of this squeezing gate, it is necessary to reduce the losses in the experimental system. The most significant optical loss is the coupling loss with OPA2, which was 11% in our experiment. This value is almost the same as the coupling loss with the optical fiber inside the module described in [19]. To reduce this coupling loss, the spatial mode of the output from OPA modules should be made closer to a perfectly circular mode. The propagation loss in the PPLN waveguide is less of a concern, as it is greatly reduced by its parametric gain. Other optical losses are caused by the variable beam splitter and the numerous mirrors and lenses placed on the lower arm in Fig. 2 to align the optical path length and the spatial mode. The former is probably due to the high loss of the cube PBS, which can be avoided by utilizing a beam splitter with a fixed splitting ratio. For the latter, the number of mirrors and lenses can be significantly reduced by tuning the spatial mode of the OPA2 output instead of tuning that of the quantum state in the lower arm. Since the output of OPA2 is classical information, it is not a problem that the loss here is increased by mirrors and lenses. Once the loss of the experimental system has been improved, the next step will be to input a coherent state into this squeezing gate. We fully expect that it will soon be possible to demonstrate ultrafast quantum teleportation and entanglement swapping with all-optical feedforward.

In our experiments, linear QIP is performed by linear all-optical measurement-device-free feedforward. For the universality of QIP, nonlinear feedforward is also necessary [29,30]. Recently, nonlinear feedforward has been achieved using FPGAs [31], but this approach will again result in bandwidth limitation by electronics. In the all-optical method, since the output of OPA2 is tolerant of optical losses to some extent, we can adopt nonlinearity by using lossy nonlinear optical devices after OPA2, such as highly nonlinear fibers, saturable absorbers, and so on. If such nonlinear optical devices require high input power, we can also use lossy amplifiers such as erbium-doped optical fiber amplifiers (EDFAs). Thus, all-optical feedforward has the potential to enable a universal, ultra-fast, and fault-tolerant quantum computer, and we have implemented the underlying technology in this experiment.

## 6.  Conclusion

Measurement-based quantum information processing (QIP) requires feedforward based on the measurement of a prepared entangled state that can be generated over the THz bandwidth. However, performing measurements and feedforward with electronics limits the operation bandwidth of quantum operations for QIP. In this paper, as an alternative to the conventional "opto-electro-optical" feedforward using a homodyne detector, we performed all-optical feedforward utilizing a PPLN waveguide as an OPA, which removes the measurement device and bandwidth limitation posed by the use of electronics. We constructed an experimental setup of an all-optical variable squeezing gate and precisely tuned the gain of the feedforward and the optical path length for broadband feedforward. We also partially canceled out the effect of dispersion due to the output fiber of the OPA module. As a result, we were able to perform all-optical measurement-free feedforward and achieve a variable squeezing gate with an ultra-broad operation bandwidth of 1.3 THz. Furthermore, since the phase locking is performed continuously in time, our all-optical squeezing gate can also be performed continuously. These results suggest that we can perform QIP at a THz clock frequency, which is improved by around four orders of magnitude. We have implemented the underlying technology for all-optical QIP and achieved a major step toward realizing an ultra-fast quantum computer.

**Appendix. Description of Parametric Amplification in a Lossy Waveguide**

In this appendix, we describe optical parametric amplification in a lossy waveguide using the quadrature operators $\hat{x}$ and $\hat{p}$. With parametric gain per unit length $g$ and optical path length $\delta z$, an ideal optical parametric amplification without loss is equivalent to a squeezing operation with a squeezing parameter of $gz$, that is $\hat{x} \to \hat{x}' = e^{-g\delta z}\hat{x}$ and $\hat{p} \to \hat{p}' = e^{g\delta z}\hat{p}$. Introducing an optical loss represented by an extinction coefficient $\alpha$, the quadrature operators $\hat{x}$ and $\hat{p}$ of light after traveling $\delta z$ are expressed as follows.

$$\hat{x}'' = e^{-g\delta z}\sqrt{e^{-2\alpha\delta z}}\hat{x} + \sqrt{1 - e^{-2\alpha\delta z}}\hat{x}_{\text{vac}}$$

$$\hat{p}'' = e^{g\delta z}\sqrt{e^{-2\alpha\delta z}}\hat{p} + \sqrt{1 - e^{-2\alpha\delta z}}\hat{p}_{\text{vac}}$$

The subscript "vac" means that they are operators for the introduced vacuum. If we prepare a waveguide of length $L = N\delta z$, the above procedure is repeated $N$ times and the output state is expressed as

$$\hat{x}_{\text{out}} = e^{-gN\delta z}\sqrt{e^{-2\alpha N\delta z}}\hat{x}_{\text{in}} + \sqrt{1 - e^{-2\alpha\delta z}}\sum_{n=1}^{N} e^{-g(n-1)\delta z}\sqrt{e^{-2\alpha(n-1)\delta z}}\,\hat{x}_{\text{vac}}^n,$$

$$\hat{p}_{\text{out}} = e^{gN\delta z}\sqrt{e^{-2\alpha N\delta z}}\hat{p}_{\text{in}} + \sqrt{1 - e^{-2\alpha\delta z}}\sum_{n=1}^{N} e^{g(n-1)\delta z}\sqrt{e^{-2\alpha(n-1)\delta z}}\,\hat{p}_{\text{vac}}^n,$$

where $\hat{x}_{\text{vac}}^n$ and $\hat{p}_{\text{vac}}^n$ represent the quadrature operators for a vacuum induced at the $n$th procedure. We can summarize the second summation term as $\sum f_n \hat{x}_{\text{vac}}^n = \sqrt{\sum |f_n|^2}\,\hat{x}_{\text{vac}}$, since none of the induced vacuum fields correlate with each other. Finally, we consider the limit $\delta z \to 0$ under the condition of $N\delta z = L$ and obtain the description of the output as follows.

$$\hat{x}_{\text{out}} = e^{-gL}\sqrt{e^{-2\alpha L}}\hat{x}_{\text{in}} + \sqrt{\int_0^L dz\, 2\alpha e^{-2(g+\alpha)z}}\,\hat{x}_{\text{vac}}$$

$$= e^{-(g+\alpha)L}\hat{x}_{\text{in}} + \sqrt{\frac{\alpha(1 - e^{-2(g+\alpha)L})}{g + \alpha}}\,\hat{x}_{\text{vac}}$$

$$\hat{p}_{\text{out}} = e^{gL}\sqrt{e^{-2\alpha L}}\hat{p}_{\text{in}} + \sqrt{\int_0^L dz\, 2\alpha e^{2(g-\alpha)z}}\,\hat{p}_{\text{vac}}$$

$$= e^{(g-\alpha)L}\hat{p}_{\text{in}} + \sqrt{\frac{\alpha(e^{2(g-\alpha)L} - 1)}{g - \alpha}}\,\hat{p}_{\text{vac}}$$

Here, we assume $1 - e^{-2\alpha\delta z} \sim 2\alpha\delta z$. From the variance of the anti-squeezed quadrature, we can calculate the efficiency $\eta$ of the optical parametric amplification with a gain per unit length $g$ and an extinction coefficient $\alpha$ as $\eta = \frac{(g-\alpha)\exp\{2(g-\alpha)L\}}{g\exp\{2(g-\alpha)L\} - \alpha}$.

**Funding.** Japan Science and Technology Agency (JPMJMS2064, JPMJPR2254); Japan Society for the Promotion of Science (22K20351, 23K13038, 23K13040).

**Acknowledgements.** The authors acknowledge supports from UTokyo Foundation and donations from Nichia Corporation of Japan. T.Y., T.N. and T.S. acknowledge financial support from The Forefront Physics and Mathematics Program to Drive Transformation (FoPM). M.E., K.T. and W.A. acknowledge supports from Research Foundation for OptoScience and Technology.

**Disclosures.** The authors declare no competing financial interests.